\documentclass[sigconf]{acmart}
\AtBeginDocument{%
  }

\acmDOI{}
\acmISBN{}
\acmYear{}
\acmConference[Tools For Thought at CHI '25]{CHI2025 Workshop on Tools for Thought}{April 26, 2025}{Yokohama, Japan}
\usepackage[multiple]{footmisc}

\usepackage{microtype}

\usepackage{titlesec}
\titlespacing*{\section} {0pt}{0.7ex}{0.5ex}
\titlespacing*{\subsubsection} {0pt}{2ex}{2ex}

\setcopyright{none}

\begin{document}

% \title{Changing Nature of Thought in the Age of AI}
% \title{Generative AI is Changing How We Think and Learn}
\title{Protecting Human Cognition in the Age of AI}
% \title{Cognitive Impacts of Generative AI and Implications for Thinking and Learning}

% \affiliation{%

\author{Anjali Singh}
\email{anjali.singh@ischool.utexas.edu}
\affiliation{%
  \institution{The University of Texas at Austin}
  \country{USA}
}

\author{Karan Taneja}
\email{ktaneja6@gatech.edu}
\affiliation{%
 \institution{Georgia Institute of Technology}
 \country{USA}
 }

\author{Zhitong Guan}
\email{klarazt@utexas.edu}
\affiliation{%
  \institution{The University of Texas at Austin}
  \country{USA}
}

\author{Avijit Ghosh}
\email{avijit@huggingface.co}
\affiliation{%
  \institution{Hugging Face, University of Connecticut}
  \country{USA}
  }


\begin{abstract}
The rapid adoption of Generative AI (GenAI) is significantly reshaping human cognition, influencing how we engage with information, think, reason, and learn. This paper synthesizes existing literature on GenAI's effects on different aspects of human cognition. Drawing on Krathwohl’s revised Bloom’s Taxonomy and Dewey’s conceptualization of reflective thought, we examine the mechanisms through which GenAI is affecting the development of different cognitive abilities. We focus on novices, such as students, who may lack both domain knowledge and an understanding of effective human-AI interaction. Accordingly, we provide implications for rethinking and designing educational experiences that foster critical thinking and deeper cognitive engagement. 
% and discuss future directions to explore the long-term cognitive effects of GenAI.
\vspace{-2mm}
\end{abstract}


% \keywords{Do, Not, Us, This, Code, Put, the, Correct, Terms, for,

\maketitle

\section{Introduction}

The many technological advancements of the 21st century, such as the internet, web search, social media, and, more recently, Generative AI (GenAI), have profoundly influenced how individuals think, learn, communicate, and work. Unlike the advent of the internet and web search, which happened at a significantly slower rate, GenAI has been adopted at a much faster rate \cite{Harvard2Generative2024, ReutersChatGPT, BIDeepSeek}. Further, with GenAI, it is now possible to synthesize vast amounts of information on the web into coherent, structured and accessible outputs. 
% Consequently, generative AI is fundamentally altering the way we think and work. 
Consequently, GenAI has the potential to fundamentally alter the way we think.
On one hand, emerging evidence highlights notable benefits of GenAI, such as increased productivity across various contexts \cite{al2024enhancing}, enhancements in the creative process for artists and designers \cite{ali2023using}, and improved learning experiences \cite{yan2024promises}. On the other hand, significant concerns have been raised about the potential long-term detrimental effects of GenAI on cognitive abilities such as critical thinking \cite{lee2025impact} and reasoning \cite{klingbeil2024trust}. Central to understanding these multifaceted impacts of AI lies the need to analyze how the nature of human thought itself is evolving. This analysis is needed to identify strategies to develop `tools for thought' in the age of AI. 

% Recently, Tankelevitch et al. \cite{tankelevitch2024metacognitive} argued that GenAI systems impose distinct metacognitive demands on users due to their inherent characteristics, such as model flexibility, generality, and originality, and that understanding these demands is pivotal for addressing the novel usability challenges posed by these systems. 
This work explores the impact of GenAI on fundamental aspects of human cognition and how other societal factors, such as heightened stress levels \cite{melchior_work_2007} and information overload \cite{woods2002can}, interact with these effects. 
% For instance, human attention spans are undergoing a notable reduction, driven by factors such as increased exposure to short-form content \cite{}. We explore how these broader societal changes can influence the use of Generative AI tools, particularly in ways that can have detrimental long-term consequences for human cognition.
While GenAI can both augment and hinder cognitive development depending on how it is used, this work primarily examines the latter and explores strategies to protect essential cognitive abilities. We focus on examining GenAI's cognitive effects on novices, such as students, rather than on experts or professionals. However, we also draw some comparisons between the two groups to better understand GenAI's multifaceted impacts.

We begin by synthesizing the existing literature on the impacts of GenAI on human cognition to delineate the various dimensions of cognition that are being influenced. Building on this, we draw upon Krathwohl's revised Bloom's Taxonomy \cite{krathwohl2002revision} and Dewey's seminal work, \textit{How We Think} \cite{dewey2022we} to explore the underlying reasons for the observed effects on human cognition. Accordingly, we propose adopting Dewey’s conceptualization of critical thinking---as reflective thinking---to both \textit{evaluate the impact} of GenAI on learners’ critical thinking and \textit{inform the design} of interventions aimed at fostering critical thinking. Further, we provide insights on applying Dewey's framework to support the evolving nature of thinking and learning.
% essential cognitive practices, such as critical thinking, particularly in educational contexts. 
% With this goal, we explore the following research questions:
% \vspace{-1mm} 
% \begin{itemize}
%     \item RQ1. \textit{What} impacts of GenAI on human cognition have been observed, based on current research?
%      \item RQ2. \textit{How} is GenAI impacting thinking and learning? 
%       \item RQ3. Accordingly, what are the implications for designing tools to support thinking and learning in the age of AI?     
% \end{itemize}
% \vspace{-2mm} 
% \section{Related Work on Impacts of Generative AI on Human Cognition}
% - reasoning/logical/decision making
% - metacognitive
% - emotional

\section{Emerging Cognitive Challenges in the AI Era}
% This section examines the current literature on how GenAI is influencing various aspects of human cognition and provides the foundation for our subsequent analysis of the underlying mechanisms and implications for educational practice.

Human cognition encompasses a range of abilities, including memory, attention, reasoning, critical thinking, and creativity. These processes enable people to acquire, process, and apply knowledge. Human cognition has been historically influenced by the integration of technology into daily life---GPS-based navigation affected spatial cognition \cite{yan2022does}, search engines impacted memory recall patterns \cite{gong2024google}, and social media and its move towards increased short-form content has contributed to reduced attention spans \cite{chen2023effect}. More recently, with the advent of GenAI, new cognitive challenges are emerging, particularly in knowledge acquisition, reasoning, learning, creativity, metacognition and critical thinking.

\subsubsection*{\textbf{Knowledge Acquisition}}
GenAI search systems, such as Perplexity.ai, SearchGPT, and Microsoft Copilot, are transforming information seeking behaviors given their ability to synthesize well-structured responses spanning multiple sources.
% , and thus are reshaping how people interact with information. 
Traditional search engines require active seeking and verification of sources \cite{shah2024envisioning}, while GenAI may shift users toward passive consumption of potentially one-sided views, diminishing their ability to discern reliable information and limiting exposure to diverse perspectives \cite{venkit2024search}. The tendency to engage with AI-generated outputs that align with pre-existing beliefs further exacerbates the echo chamber effect \cite{cinelli2021echo}. Sharma et al. \cite{sharma_generative_2024} show that people interacting with a GenAI search system placed selective attention or retention on GenAI responses and spent more time on consonant versus dissonant opinions. Further, the populistic approach of GenAI search systems-- prioritizing widely accepted perspectives while sidelining alternative views-- can contribute to homogenization, marginalize less visible but valuable information and suppress marginalized identities, languages, cultural practices, and epistemologies \cite{solaiman2023evaluating, venkit2024search}.

\subsubsection*{\textbf{Reasoning}}
Recent studies highlight how GenAI can influence decision-making, user trust, and susceptibility to misinformation.
A survey conducted with 285 university students revealed that they perceive increased AI usage to result in poorer decision-making and an increase in laziness \cite{ahmad2023impact}.
Klingbeil et al. \cite{klingbeil2024trust} found that individuals tend to over-rely on AI-generated advice, even when it contradicts their own reasoning or available contextual information. Alarmingly, the mere knowledge of advice being generated by an AI can cause people to over-rely on it. AI-generated explanations, particularly those that are deceptive, have also been shown to significantly influence users’ beliefs by amplifying belief in false news headlines and undermining truthful ones \cite{danry2024deceptive}. 
% \vspace{-4mm}
\subsubsection*{\textbf{Learning}}
While GenAI can facilitate the efficient acquisition of information on a given topic, it may not necessarily support effective learning or knowledge retention. For instance, a study involving a writing task found that while ChatGPT can significantly improve short-term task performance, it does not boost knowledge gain and transfer \cite{fan2024beware}. Concerns regarding skill atrophy \cite{niloy2024chatgpt} and over-reliance on GenAI \cite{song2023enhancing} have also been raised in educational contexts. 
In programming education, GenAI support has been shown to hinder students’ development of problem-solving skills, particularly among novice learners. For instance, support from AI code-generators \cite{kazemitabaar2023novices} or AI-generated feedback \cite{pankiewicz2023large} has been found to make learners over-reliant on such support. Further, students struggle to perform as well in the absence of AI assistance \cite{darvishi2024impact}. Furthermore, students often overestimate their learning gains from AI tools, mistaking ease of task completion for genuine understanding \cite{lehmann2024ai}. Similarly, while AI-assisted tutoring in math education led to improvements in short-term performance, students who relied on AI struggled more when AI access was removed, suggesting a potential detriment to long-term learning outcomes \cite{bastani2024generative}.
% \vspace{-4mm}
\subsubsection*{\textbf{Creativity}}
GenAI is reshaping our understanding of creativity, both in how we define it and how we express it. Content generated using GenAI can be more creative or less so, depending on how we define creativity. 
% An experiment conduced with 600 university students~\cite{niloy2024chatgpt} found a negative correlation between ChatGPT use and writing creativity in terms of content accuracy and originality, but a positive correlation with elaboration and content presentation. 
% In another study, 
For instance, Doshi and Hauser \citep{DoshiHauser2024} found that less creative writers could produce more engaging and original stories with GenAI. However, this improvement came at the cost of collective diversity, as stories generated with AI assistance were more similar to each other than human-written ones. 
% A possible explanation for this homogeneity lies in the data-driven nature of generative AI models. 
As Peschl \cite{Peschl2024} argues, these models risk stifling creativity by recycling existing knowledge, creating a feedback loop that reinforces repetitive patterns in both ideas and user interactions. 
% While LLMs can enhance fluency, flexibility, and elaboration \cite{niloy2024chatgpt}, they facilitate rapid enumeration of possibilities rather than development of truly original ideas, and they make users feel less responsible for the ideas they generate \cite{anderson2024homogenization}.
Taking a more nuanced approach to understanding GenAI's impact on creativity, Kumar et al. \citep{kumar2024human} 
% examined the impact of GenAI by distinguishing between convergent and divergent thinking in the idea generation process. They 
found that LLMs are more effective for convergent thinking (goal-oriented tasks) while they hinder divergent thinking, which requires exploration and unconventional approaches. 
% This suggests that heavy reliance on LLMs may lead to more homogenized ideas, as users converge on similar solutions.
Sternberg's work \cite{Sternberg2024DoNW} views creativity not as merely a cognitive process or hidden information, but as a skill. It argues that widespread use of GenAI can lead to a decline in human creativity, as people reduce exercising their own creative abilities. 
% More concerning, he argues, is the risk that we may begin to accept AI-generated outputs of middling creativity as our own, ultimately lowering our standards and expectations.

\subsubsection*{\textbf{Metacognition and Critical Thinking}}
Metacognition refers to the awareness and regulation of thinking \cite{winne2017cognition}. It improves critical thinking by enabling reflection, assessment, and adjustment of thought processes for better learning.
GenAI systems not only impose metacognitive demands on their users \cite{tankelevitch2024metacognitive} but also influence how people regulate their cognitive abilities, including critical thinking \cite{lee2025impact}.
With knowledge workers, a recent survey suggests that greater confidence in GenAI is linked to reduced critical thinking \cite{lee2025impact}. In educational contexts, the widespread use of GenAI tools has sparked concerns about `cognitive offloading' \cite{risko2016cognitive}, i.e., when students delegate cognitive tasks to AI, reducing their own cognitive engagement which impacts their ability to self-regulate and critically engage with learning material \cite{fan2024beware}. 
% Studies have shown that frequent exposure to AI-generated responses can lead to what researchers term "metacognitive laziness" \cite{fan2024beware}, where individuals increasingly offload cognitive tasks to AI tools. This behavioral shift can result in decreased internal cognitive engagement over time, ultimately affecting learners' ability to self-regulate and critically engage with learning material. The phenomenon is particularly evident in educational settings, where students may prematurely resort to AI assistance before attempting to reason through problems independently. 
Another study found a significant negative correlation between frequent AI tool usage and critical thinking abilities, with cognitive offloading as a mediating factor \cite{gerlich2025}. Further, younger participants were found to exhibit greater reliance on AI tools. Prather et al. \cite{prather2024widening} found that GenAI tools can compound the metacognitive difficulties novices face when learning to program, especially for low-performing students.
This work, along with other recent work on novices learning with GenAI tools \cite{margulieux2024self}, highlights the need to explicitly teach metacognitive skills to novice learners.

% \subsection{Generative AI and Human Reasoning (New)}

% Moreover, the integration of AI into daily cognitive tasks has introduced new dimensions to human reasoning processes. Research by Gajos et al. \cite{} suggests that even when AI systems provide explanations or citations, users often fail to engage in deeper cognitive processing of this content. This pattern of shallow engagement raises concerns about the long-term impact on human reasoning capabilities, particularly in contexts requiring nuanced understanding or critical evaluation.

% \vspace{-4mm}
\section{Mechanisms of AI-Induced Cognitive Change}
To understand how GenAI affects human cognition, we draw upon Bloom's Revised Taxonomy and Dewey's conceptualization of reflective thinking. These foundational frameworks provide complementary perspectives on thinking and learning processes and help illuminate the mechanisms through which AI influences cognitive development. Further, we discuss how other societal factors such as widespread stress and anxiety interact with the effects of GenAI on human cognition.

\subsubsection*{\textbf{Bloom's Revised Taxonomy}}
Towards providing a framework for categorizing educational objectives, Bloom's Revised Taxonomy \cite{krathwohl2002revision} comprises two dimensions: knowledge and cognitive processes. The knowledge dimension encompasses factual, conceptual, procedural, and metacognitive knowledge. Cognitive processes, in increasing order of complexity, are to remember, understand, apply, analyze, evaluate, and create. 
Learning involves a complex interplay of acquiring different types of knowledge, engaging in cognitive processes, and utilizing metacognitive knowledge to enhance learning. The process can be broken down into: 
\begin{enumerate}
    \item Gradual acquisition of various types of knowledge, including declarative, procedural, and schematic knowledge, which form the foundation of further cognitive processing \cite{winne2021cognition}.
    \item Engagement in cognitive processes to transform this knowledge into deeper understanding. This involves applying cognitive operations to experiences, and is affected by prior knowledge. Misconceptions (such as due to misinformation) or indiscriminate application of skills impedes learning \cite{drigas2017consciousness}.
    \item Metacognitive engagement, which involves acquiring meta-cognitive knowledge, including awareness and regulation of one's cognitive processes. It helps learners monitor and control their learning strategies, making adjustments as needed. A challenge learners face is overcoming overconfidence about what they know \cite{winne2021cognition}.
    % \vspace{-2mm}
\end{enumerate}

\begin{figure}[t]
  \centering
  \includegraphics[width=\columnwidth]{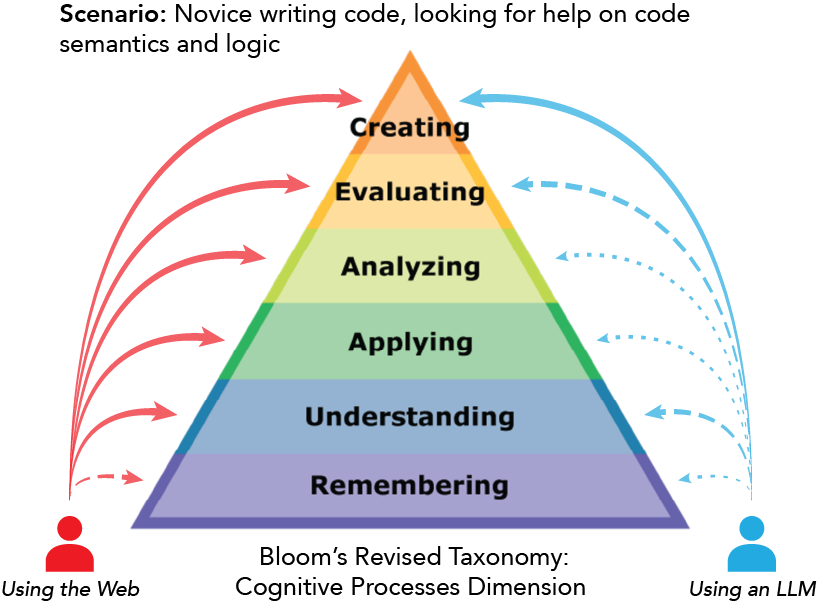}
  \caption{Comparing cognitive processes engaged in the given scenario when a novice uses the web versus an LLM for support, with fainter arrows indicating weaker engagement.}
  \vspace{-4mm}
    \label{fig:skips}
\end{figure}

AI is influencing how the two dimensions of Bloom's Revised Taxonomy interact with each other. GenAI accelerates access to factual and procedural knowledge. In contrast, traditional learning involved gradual knowledge acquisition through books, interactions with teachers and peers and, more recently, web search. With GenAI, learners can access instantly synthesized and complex information. However, this acceleration may bypass important cognitive processes that typically occur during slower, deliberate learning, as evidenced in recent studies \cite{bastani2024generative, darvishi2024impact, fan2024beware, gerlich2025, lehmann2024ai, risko2016cognitive}. For example, consider a novice programmer working on a programming problem and searching for guidance on code semantics and logic (Figure \ref{fig:skips}). In the absence of LLMs, the novice would skim through various online resources to acquire knowledge. In this process, they will need to expend effort to \textit{understand} information from each source, \textit{apply} this information to the problem at hand, \textit{analyze} various approaches by synthesizing ideas from different sources, and \textit{evaluate} different solutions along with their underlying sources to \textit{create} a working solution.
% \textit{evaluating} information from different sources,  and \textit{applying} it to the problem at hand. 
However, when using LLMs in a similar scenario, they may miss opportunities to develop and practice these essential cognitive skills---particularly remembering, applying, analyzing, and, sometimes, evaluating---by outsourcing them to LLMs.
This, in turn, impedes the development of metacognitive skills, which are acquired through regular practice and assessment of different cognitive processes. Although the specific cognitive processes impacted by GenAI will vary depending on the context \cite{lee2025impact}, this example serves as an illustrative case.

Experiences of cognitive difficulty prompt more analytical reasoning, as learners are more likely to engage in deliberate thought when their intuition is challenged \cite{alter2007overcoming}. Overreliance on GenAI can reduce such cognitive difficulty, which can reduce the activation of deeper metacognitive processes necessary for analytical reasoning.
This frequent outsourcing of essential cognitive tasks to AI by learners rather than developing their own metacognitive strategies has also been referred to as ``metacognitive laziness'' \cite{fan2024beware}.
Younger or novice learners can be particularly susceptible to metacognitive laziness \cite{kazemitabaar2023novices, gerlich2025, pankiewicz2023large} due to their limited experience in employing diverse cognitive strategies, and, consequently, a weaker understanding of effective learning practices. 

In contrast to novices, Imundo et al. argue that experts are better positioned to leverage GenAI for high-level decision-making, as they possess well-structured content and procedural knowledge within their domain \cite{imundo2024expert}. This expertise allows them to formulate effective prompts that guide GenAI in producing relevant responses, by drawing on accurate domain-specific terminology and knowledge structures. Therefore, GenAI has the potential to enhance expert cognition, particularly as a collaborative tool for offloading lower-level cognitive tasks \cite{imundo2024expert}, fostering creativity through iterative questioning \cite{beghetto2023new}, improving text comprehension \cite{yeoacademic}, and facilitating learning through approaches such as the Socratic method \cite{yang2005using}. However, even for experts, over-reliance on AI poses risks, as it could hinder the deliberate practice necessary for refining and sustaining high levels of expertise \cite{ericsson2006influence}.

\subsubsection*{\textbf{Dewey's Theory of Reflective Thinking}}

% \begin{quote}
%     % \centering
%     ``We do not learn from experience...we learn from reflecting on experience'' -- John Dewey
% \end{quote}
In `How We Think' \cite{dewey2022we}, Dewey identifies four types of thought, ranging from the broadest to the most disciplined--(i) mere awareness, (ii) imaginative thought, which excludes sensory experience, (iii) belief based on evidence (whether examined or not), and (iv) reflective thought involving conscious evaluation of evidence. Dewey argues that reflective thought is most essential for deep and meaningful learning, and describes several prerequisites for it:
\begin{itemize}
    \item A state of perplexity, confusion or doubt prompting inquiry
    \item Prior experience and knowledge to draw upon
    \item Active, persistent consideration of ideas
    \item Suspended judgment during further inquiry and a tolerance for uncertainty, or ``\textit{mental unrest and disturbance}'' 
    \item The ability to connect and evaluate related ideas
\end{itemize}

GenAI can potentially disrupt these prerequisites in several ways. First, beyond the initial state of perplexity, confusion or doubt that prompts an individual to initiate interaction with a GenAI system, the \textit{immediate, synthesized} GenAI responses may diminish subsequent moments of cognitive dissonance which are necessary for initiating further reflective thinking. Secondly, GenAI's \textit{tendency to align with preexisting beliefs} \cite{venkit2024search, sharma_generative_2024} and \textit{displaying human-like social desirability biases} \cite{salecha2024large} can reinforce existing perspectives rather than challenge them, thereby undermining the active and persistent engagement with diverse viewpoints necessary for reflective thinking. 
Next, responses provided by LLMs tend to be \textit{confident}, lacking explicit representations of uncertainty or the ability to communicate their absence \cite{kidd2023ai}. This increases the \textit{persuasiveness} of AI-generated content, which can discourage the suspension of judgment that is fundamental to deep reflection. Even when AI-generated responses include explanations, people may not cognitively engage in deeper processing of the provided content and explanations \cite{gajos2022people}. This can result in a failure to detect AI misinterpretations or inaccuracies, leading to inaccurate beliefs, and reinforcement of biased views.
Finally, the \textit{structured and coherent nature} of synthesized AI responses can create an illusion of comprehensive understanding, when in reality, people may only achieve a superficial grasp of the underlying topic or concept. This phenomenon is supported by recent work demonstrating that learners frequently overestimate their learning from AI tools \cite{lehmann2024ai}. 
\vspace{-1mm}
\subsubsection*{\textbf{Societal Factors Interacting with AI's Cognitive Impacts}}
Beyond AI, various societal factors interact with GenAI's cognitive impacts. The accelerated pace of modern life has contributed to widespread chronic stress and anxiety \cite{melchior_work_2007}, which are known to impair cognitive functioning \cite{mcewen_stress_1995}. Taken together with GenAI’s ability to simulate human-like behavior--such as empathy, memory of past conversations, and humor--this can lead to emotional dependence on AI \cite{mit2025emotion, zhou_design_2020}. Such misplaced trust in AI can have severe consequences \cite{vice2023suicide,guardian2024mother}, as strong emotional responses can interfere with reasoning and reflective thinking \cite{pham2007emotion}. 
% Heightened stress levels have also been linked to an increase in unethical behavior \cite{kouchaki_anxious_2015}, exacerbating concerns about AI’s influence on cognitive and moral decision-making.
These issues intersect with information overload--the issue of exponential growth of available information that outpaces our ability to interpret it \cite{woods2002can}. As Herbert Simon argued, attention becomes the limiting resource in such information abundance \cite{simon2013administrative}. While increased access to data enhances decision-making in principle, the overwhelming volume of information complicates the identification of what is most relevant and meaningful. In today’s digital landscape---including search engines, social media, and now GenAI---human attention spans and regulatory capacities are increasingly strained \cite{ralph2014media}. This, in turn, impairs our ability to critically engage with and derive meaningful insights from GenAI generated content.
% Finally, decreasing attention spans, exaggerated by social media's move towards short form content, can further inhibit GenAI users' ability to engage deeply with the generated content, further impairing decision making and learning. 

% Finally, the proliferation of AI-generated content on social media \cite{Georgetown2024ai-generated}, especially short-form content \cite{chen2023effect}, can further diminish people's attention spans as well as ability to regulate attention \cite{zimmerman2023attention}, which has been associated with diminished cognitive processing \cite{xanidis2016association} and poorer learning outcomes \cite{kokocc2021mediating}.
% Finally, technology use and the COVID-19 pandemic have significantly reduced in-person social interactions \cite{}, negatively impacting students’ learning experiences \cite{}. The Community of Inquiry framework emphasizes the critical role of the social environment in learning outcomes \cite{}. Yet, GenAI can further contribute to this issue by giving students an illusion of learning and appearing more knowledgeable than their peers or teachers \cite{}.

% \vspace{-2mm}

\section{Supporting Thinking \& Learning in the AI Era}
Based on our analysis of GenAI's cognitive impacts, we now provide implications to support thinking and learning in the AI era. 

\subsubsection*{\textbf{Implications for Educators and Test Designers}}
Using Bloom’s Revised Taxonomy, we explored how the cognitive processes that support learning are being impacted in contexts mediated by GenAI. Our analysis reveals that the evolving nature of thinking and learnin-g---characterized by increased human-AI interactions---necessitates a greater emphasis on teaching critical and evaluative skills to ensure effective engagement with GenAI tools. However, current curricula and high-stakes tests prioritize fostering skills at which AI excels, such as formulaic decision-making \cite{cao2023navigating}. To address this, test designers should consider emphasizing critical and evaluative skills in institutional and standardized tests, as these directly inform teaching practice \cite{pedulla2003perceived}. Developing learning activities that require students to actively critique GenAI outputs \cite{singh2024bridging, singh2024empowering, oates2025chatgpt} can also be helpful for fostering the development of such skills. 

At the beginning of learning a new skill, cognitive effort and persistence are essential--- also conceptualized as `productive struggle' \cite{stanford2025productive} or `productive failure' \cite{kapur_productive_2008}. Similarly, Brown et al. \cite{brown2024cognitive} posit building `cognitive endurance' among students, \textit{i.e.}, the ability to sustain effortful mental activity over a continuous stretch of time. This effort, even when it is challenging or frustrating, is needed for deep understanding. 
Therefore, in the early stages of learning, AI use should be minimal, primarily serving functions such as providing formative feedback \cite{irons2021enhancing}. This can be implemented using guardrails in educational AI tools that facilitate a gradual increase in learner-AI interactions, ensuring that learners have the ability to exercise independent judgment when they seek AI assistance. 
% The optimal timing and manner of integrating AI tools to support student learning within specific educational contexts remains an open question, necessitating further research.
% During this initial phase of learning, AI use should be very limited, such as for providing formative feedback. This could be achieved by embedding guardrails in AI-powered educational tools, wherein learners' active engagement with AI gradually increases over time, to ensure that they are able to exercise their independent judgement even when seeking help from AI. Moreover, principles of cognitive effort and persistence should be integrated into AI literacy education for students and educators, to emphasize their necessity and promote effective use of GenAI.
% AI literacy is also essential to ensure responsible use of AI for learning, especially for younger learners.

\subsubsection*{\textbf{Promoting Active Engagement with GenAI Tools}}
GenAI influences cognitive development in complex and multifaceted ways, with its impact largely dependent on who engages with the technology and how it is used. In particular, it is passive engagement and over-reliance on AI that pose significant risks to users’ thinking. This calls for the development of novel evaluation frameworks to assess learning, particularly in educational contexts---moving beyond traditional metrics to assess dimensions such as cognitive and metacognitive engagement, critical thinking, and depth of learning.
We propose using Dewey’s prerequisites for reflective thinking as a framework for both: (i) \textit{identifying passive AI use} based on unmet prerequisites of reflective thinking in human-AI interactions, and (ii) \textit{designing interventions} to foster critical thinking. We now illustrate several applications of this framework.

The first prerequisite---a state of perplexity, confusion, or doubt prompting inquiry---requires learners to reflect on gaps in their understanding, a process that may be undermined by the immediacy of AI-generated responses. 
To support this pre-requisite, one approach is to enable learners to highlight parts of GenAI responses. This could also provide feedback to the GenAI system, enabling it to generate more relevant responses to support a learner's thinking. Similarly, introducing small amounts of forced yet desirable engagement with AI-generated outputs---conceptualized as `friction' in human-AI interaction \cite{kazemitabaar2024exploring}---before learners can use the generated outputs can help promote more active engagament.

Next, GenAI chatbots may prioritize surface-level discussions over persistent exploration of ideas, which is a another prerequisite of reflective thinking. This tendency may be especially pronounced for novices, who often lack the skills for effective prompt engineering and a foundational understanding of the subject matter \cite{federiakin2024prompt}. 
% Additionally, these chatbots do not account for learners’ prior knowledge.
Building on schema theory \cite{derry1996cognitive}, which describes how knowledge is organized in the brain as interconnected frameworks or `schemas', integrating cognitive schemas \cite{andersonpearson1984} into educational tools can facilitate learners' ability to identify connections between related ideas and link them to prior knowledge, thereby fostering comprehensive understanding \cite{Rusen2024}. 
% This can help learners connect new information with their prior knowledge.
For such schema-based learning, an example tool could have an interactive interface with a navigable knowledge graph along with a conversational AI agent. When encountering new information, the tool could dynamically highlight relevant schemas in the knowledge graph to activate prior knowledge. 

% To encourage suspended judgment, a prerequisite for reflective thinking that is affected by GenAI's persuasiveness and tendency to align with preexisting beliefs, a promising approach is the use of metacognitive prompts
For suspended judgment---a prerequisite for reflective thinking undermined by GenAI’s persuasiveness and alignment with users’ preexisting beliefs---metacognitive prompts offer a promising intervention
% \footnote{A tool we are developing to guide learners' interactions with GenAI search systems using metacognitive prompts.}
\cite{lin1999supporting}. These prompts nudge learners to pause and reflect at critical moments, for instance, by encouraging them to consider alternative perspectives or assess their comprehension of GenAI outputs. Metacognitive prompts can help learners be more aware of their thinking, as even those who possess metacognitive knowledge and skills may not always know how to use them spontaneously \cite{bannert2013scaffolding}. Similarly, recent work has explored the concept of AI-generated `provocations', which highlight the risks, biases, limitations, and alternatives to GenAI recommendations
% , to encourage critical engagement rather than passive acceptance of AI suggestions
\cite{drosos2025makes, sarkar2024ai}. 
% Additionally, Dewey emphasized that training in thinking should not merely establish the prerequisites for reflective thought but should actively direct and utilize a learner's existing cognitive abilities. 

\section{Conclusion}
This paper examines the impacts of GenAI on human cognition, providing several key insights that have important implications for education, interaction design, and cognitive development in the AI era. 
Much of the current research on AI's cognitive impacts relies on short-term studies, making it difficult to draw definitive conclusions about long-term longitudinal effects. This presents a pressing need for long-term studies on how sustained AI use affects cognitive development, particularly for younger users who are most vulnerable to the negative effects of GenAI overuse. 

\bibliographystyle{acm}
\bibliography{REFERENCES,karan-zotero}

\end{document}